# ROSAT PSPC spectra of six PG quasars and PHL 1657

**Jörg P. Rachen**[1], **Karl Mannheim**[2], and **Peter L. Biermann**[1]

[1] Max-Planck-Institut für Radioastronomie, Auf dem Hügel 69, D-53121 Bonn, Germany
[2] Universitäts-Sternwarte, Geismarlandstr. 11, D-37083 Göttingen, Germany

**Abstract.** We report results from the spectral analysis of pointed *ROSAT* PSPC observations of six PG quasars and the weak-bump quasar PHL 1657. Disregarding other frequency bands, the PSPC data are represented best by simple power-law source spectra ($dN/dE \propto E^\Gamma$) with a slope $\Gamma = -2.5 \pm 0.4$ and do not show evidence for a more complex structure. However, within the limits given by statistics and systematical errors, a superposition of two power-law spectra, which allows a connection to the observed UV fluxes and to the generally flatter hard-X-ray spectra observed by *EXOSAT* and *Ginga*, is found to be indistinguishable from a simple power-law in the PSPC energy band. For the PG quasars, a direct connection to the observed UV flux is obtained with a steep soft-X-ray slope $\Gamma_{\rm sx} = -3.1 \pm 0.3$, fixing the hard-X-ray slope. For PHL 1657, the soft-X-ray spectrum seen by *ROSAT* is much steeper than the *Ginga* spectrum, rising towards the EUV in $\nu F_\nu$, and it can not be extrapolated down to meet the exceptionally weak UV flux. A Wien-shaped thermal UV/soft-X-ray bump with a temperature of $\sim 50\,\mathrm{eV}$ connects the UV and soft-X-ray spectra in all cases, but gives only a poor representation of the high signal-to-noise PSPC spectra. We discuss implications of our results for models of the X-ray emission in quasars.

Three more pointed PSPC observations, covering the radio galaxies 3C 433, 3C 83.1 and the quasar PG 2214+139 (Mkn 304), did not yield enough counts for a spectral analysis. For these sources, and four high redshift quasars from the Hewitt & Burbidge catalogue located in the observation fields, we present brief results or limits for their soft-X-ray fluxes.

**Key words:** quasars: general – quasars: individual PHL 1657 – X-rays: general

*Send offprint requests to*: J. Rachen (jrachen@mpifr-bonn.mpg.de)

## 1. Introduction

Quasars and other Active Galactic Nuclei (AGN) are known to emit electromagnetic radiation and particles over a very broad energy band. For the majority of quasars, however, the most rapid variability and the bulk of the emitted electromagnetic luminosity are observed in the optical/UV/soft-X-ray spectral domain, the so-called big blue bump. Therefore it is commonly believed that this spectral component holds the key for unveiling the nature of these enigmatic objects.

An appealing explanation for the origin of the big blue bump is accretion through a Keplerian disk (Shakura & Sunyaev 1973, Laor & Netzer 1989) or a non-Keplerian torus (Madau 1988) onto a supermassive black hole (Rees 1984). The accretion disk scenario derives its attractiveness mainly from the knowledge that accretion disks seem to occur frequently in stellar systems like close binaries or protostars. Moreover, accretion in stellar systems is often accompanied by collimated bipolar outflows resembling the radio jets of AGN.

One of the strongest implications of $\alpha$-disk models is a radial temperature scaling which is difficult to reconcile with the results of recent multiwavelength monitoring campaigns (Clavel et al. 1992). Moreover, the rapid soft-X-ray variability (Boller et al. 1993, Molendi & Maccacaro 1994) and the ubiquity of soft excesses (Walter & Fink 1993; Walter et al. 1994) found in AGN are suggestive of a hard tail of the big-blue-bump emission component reaching into the soft-X-ray range. This is expected for accretion disks radiating with near-Eddington accretion rates (Czerny & Elvis 1987, Ross et al. 1992). However, results of extensive accretion disk model fitting of AGN multifrequency data are consistent with *cold* sub-Eddington disks with the thermal emission peaking in the UV (Sun & Malkan 1989, Wandel 1991). On the other hand, observations of some highly redshifted quasars do not show a significant break in the UV spectrum consistent with a source-frame big blue bump reaching into the EUV band (Bechtold et al. 1984, Reimers et al. 1992). If the quasar luminosity is proportional to the Eddington luminosity, the effective temperature decreases with luminosity as $T_{\rm eff} \propto L^{-\frac{1}{4}}$. Thus, naively, thermal soft-X-ray excesses are not expected for the most luminous quasars. Unfortunately, the most luminous quasars have the greatest redshifts, so that the source-frame soft-X-ray band shifts to the EUV and this crucial test becomes difficult (Saxton et al. 1993).

Another approach to the problem of the origin of AGN soft-X-ray spectra is motivated from their hard-X-ray properties. Reprocessing of hard X-rays by cold, optically thick matter pro-duces a thermal big blue bump and a Compton-reflection hump at high energies (Lightman & White 1988, Guilbert & Rees 1988). Currently, it seems that the non-detection of radio-quiet AGN in $\gamma$-rays (Maisack et al. 1995) strongly argues against this scenario on energetical grounds. If, however, the prime energizer is a gas of relativistic protons rather than nonthermal photons, this objection is removed.

While steep soft-X-ray spectra seem to belong to a broad thermal emission feature, hard X-rays could be of thermal or nonthermal nature. Radio-loud AGN emit $\gamma$-rays and show evidence for Doppler-boosted X-ray emission from a jet, e.g. due to the synchrotron-self-Compton mechanism (Maraschi et al. 1992), making a clear case for nonthermal emission. On the other hand, radio-quiet AGN have steeper X-ray spectra and do not show evidence for $\gamma$-rays, which could indicate thermal Comptonization as the origin of the hard X-rays (e.g., Haardt & Maraschi 1991). Remarkably, the hard-X-ray slopes $\Gamma_{\rm hx} \sim -2$ for radio-quiet and $\Gamma_{\rm hx} \sim -1.5$ for radio-loud AGN (Wilkes & Elvis 1987) are the generic values for pair cascades in a high and low radiation compactness region, respectively (Zdziarski & Lightman 1985). In the high compactness region close to the UV-emitting accreting flow, a kinematic anisotropy directs most $\gamma$-ray production towards the source of the UV photons (Mannheim 1995), and in the low compactness region of a relativistic jet, Doppler-boosting and strong magnetic fields lead to unsaturated synchrotron cascades (Mannheim & Biermann 1992). Pair cascades could also arise if an accretion disk is irradiated by energetic protons (Niemeyer 1995).

Attempts to discern the shape of the intrinsic X-ray spectrum are hampered by 'warm' gas in the line-of-sight. This gas with intermediate ionization leads to atomic absorption and line emission features imprinted on the observed X-ray spectrum. It depends on the geometry of the emission region (e.g., filling and covering factors) to what extent reflection, absorption and emission contribute to the observed total spectrum (Netzer 1993, Reynolds & Fabian 1995). Steep soft-X-ray spectra can result from unresolved absorption edges and X-ray line emission (Ross & Fabian 1993, Czerny & Życki 1994). Reflection and fluorescence by ambient gas flatten the observed hard X-ray spectrum and lead to iron line emission. Therefore, it seems that the more luminous quasars are less affected by reprocessing than Seyferts, which generally show flatter hard-X-ray spectra with $\Gamma_{\rm hx} \sim -1.7$ and iron fluorescence.

The transition regime from the big-blue-bump emission component to the hard X-rays is particularly important to improve the theoretical understanding of quasars. In this paper we therefore investigate the ROSAT X-ray spectra of a sample of quasars. In Sect. 2 we introduce the observed quasar sample and describe the data analysis thoroughly. Finally, in Sect. 3 various model approaches are discussed critically in the light of the observational results.

For compatibility with other papers, a Hubble constant of $H_{\rm o} = 50 \, {\rm km \, s^{-1} \, Mpc^{-1}}$ and a cosmological deceleration parameter $q_{\rm o} = \frac{1}{2}$ are adopted throughout. Spectral slopes are generally given for photon number spectra, $dN/dE \equiv f_E \propto E^\Gamma$; the relation to the commonly used energy flux spectral index is $\Gamma = -(\alpha_E + 1)$. We refer to results of earlier observations of our sources, and of PG quasars in general, obtained by the X-ray satellite observatories Einstein, EXOSAT and Ginga. The sensitivity range of their imaging X-ray proportional counters IPC, LE/ME and LAC, respectively, are shown in comparison to the ROSAT PSPC in Fig. 2, in Section 2.4.

## 2. ROSAT observations

### 2.1. Description of the sample

We observed sources drawn from the optically complete sample of PG quasars, supplemented by the quasar PHL 1657. In addition, we briefly report our observations of two sources from the complete sample of 3CRR FR-I radio galaxies.

The scientific motivation for investigating PG quasars was to check on the physics of the central source, and to see, among other things, whether the X-rays could power the FIR, which had been found to be thermal (Chini et al. 1989). In our sample the ratio of FIR to X-ray luminosity is typically much larger than unity; so we immediately know that the X-rays cannot easily power the FIR, since we have evidence at the same time from the spectrum that absorption is not strong (i.e. does not reach several keV range). As a counter-example to the big-bump PG quasars we wanted to investigate a weak-bump quasar, to check for a correlation between UV and soft-X-ray properties.

The random-selected PG quasar subsample investigated by us consists of the following sources:

*PG 0844+349, PG 1411+442* and *PG 2214+139*: Low-redshift ($z < 0.1$), radio-quiet quasars with disk luminosities near to $10^{45}$ erg s$^{-1}$. PG 1411+442 shows broad absorption lines and is located in a galaxy merger.

*PG 0026+129* and *PG 0953+414*: Intermediate-redshift ($0.1 < z < 0.3$), radio-quiet quasars with disk luminosities of $2 \cdot 10^{45}$ and $10^{46}$ erg s$^{-1}$, respectively.

*PG 1302–102*: Intermediate-redshift flat-spectrum radio quasar (PKS 1302−102), which has a disk luminosity of $10^{46}$ erg s$^{-1}$.

*PG 1634+706*: High-redshift ($z = 1.334$) radio-quiet quasar; the extreme disk luminosity of almost $10^{48}$ erg s$^{-1}$ requires a black hole mass of more than $10^{10}$ solar masses even for maximal accretion.

*PHL 1657 (PKS 2135–147)*: Intermediate redshift radio-loud weak-bump quasar; this source is a paradigm for quasars where the optical/UV spectrum is much steeper than for PG quasars (McDowell et al. 1989). Its FR-II radio structure and its large ratio of extended to core flux indicates a rather large inclination of the jet axis with respect to the observer (Morganti et al. 1993).

Disk luminosities express integrated big-blue-bump luminosities in the accretion disk picture (Falcke et al. 1995b). All radio-quiet PG quasars in the sample are radio detected with fluxes of order 1 mJy (Kellermann et al. 1989), thus "radio-quiet" should not be misunderstood as "no radio emission". Except PG 0953+414 and PG 1411+442, the PG quasars are Einstein detected. hard-X-ray (2–10 keV) spectral indices are given for PG 0026+129 ($\Gamma_{\rm hx} = -1.86 \pm 0.39$, *EXOSAT*, Comastri et al. 1992) and PHL 1657 ($\Gamma_{\rm hx} = -1.81 \pm 0.06$, *Ginga*, Williams et al. 1992).

**Table 1.** *ROSAT* PSPC observation record

| Field # | observation date | pointing position (J2000.0) RA | DEC | # OBIs | total accepted time seconds |
|---|---|---|---|---|---|
| 1 | 1991, March 15,16 | $16^h34^m28.8^s$ | $+70°31'48.0''$ | 5 | 9357 |
| 2 | 1991, April 10,11,20 | $08^h47^m43.2^s$ | $+34°45'00.0''$ | 3 | 4450 |
| 3 | 1991, June 26,27 | $14^h13^m22.6^s$ | $+44°00'00.0''$ | 2 | 25319 |
| 4 | 1992, Feb 27 | $03^h18^m16.7^s$ | $+41°51'36.0''$ | 1 | 2531 |
| 5 | 1992, April 30 | $09^h56^m52.7^s$ | $+41°15'36.0''$ | 2 | 7070 |
| 6 | 1992, May 19 | $21^h23^m45.5^s$ | $+25°04'12.0''$ | 1 | 838 |
| 7 | 1992, May 26 | $22^h17^m12.0^s$ | $+14°14'24.0''$ | 4 | 7409 |
| 8 | 1992, July 19,20 | $13^h05^m33.5^s$ | $-10°33'36.0''$ | 2 | 3168 |
| 9 | 1993, June 23,24 | $00^h29^m14.3^s$ | $+13°16'12.0''$ | 2 | 2727 |
| 10 | 1993, Nov 15 | $21^h37^m45.5^s$ | $-14°33'00.0''$ | 1 | 3249 |

In the case of the 3CRR FR-I radio galaxies the scientific argument was to check on the unification scheme for BL Lac sources and Fanaroff-Riley class I radio galaxies (Fanaroff & Riley 1974). Clearly, the weak detections of the two sources actually observed by us, *3C83.1* and *3C433*, do not allow a detailed physical discussion, but the results obtained for their total X-ray luminosities may be of interest for unified AGN models and are thus included in our presentation.

Additionally, the high-redshift quasars 0843+349, 1304–107, 2134–149 and 2135–145 from the Hewitt & Burbidge catalogue (1991, hereafter referred to as HB) were located in the observation fields, and we include them in our data analysis for further reference.

### 2.2. Observations and data analysis

The sample objects were observed with the *ROSAT* (Trümper 1983) PSPC-B (Position Sensitive Proportional Counter, Pfeffermann et al. 1987), in the pointing mode between March 1991 and November 1993. Table 1 gives pointing position, observation date, total accepted time and the number of observation intervals (OBIs) of the 10 observation fields. For the 1991 observations we used the reprocessed data sets released by the *ROSAT* Scientific Data Center (RSDC) after February 1994, but we did not find substantial changes to the original data sets. Data were processed using the Extended Scientific Analysis System (EXSAS, Zimmermann et al. 1994), developed at the Max-Planck-Institut für Extraterrestrische Physik.

Sources were identified using the EXSAS maximum likelihood algorithm (DETECT/MAXLIK), applied to the PSPC image using two different source position lists: One produced directly from the PSPC image by the EXSAS sliding window source detection algorithms DETECT/LOCAL and DETECT/MAP, and the other taken from the NASA/IPAC Extragalactic Database (NED). The PG quasars of our sample and HB 0843+349 could be identified by the image source detection with a confidence level of more than 10-$\sigma$, 3C83.1 was found with a confidence level of about 5-$\sigma$, all within the PSPC image resolution ($\sim 15''$) from their catalogue positions. 3C433, and the quasars HB 2134−149 and HB 2135−145 were not found by the source detection technique, but at their catalogue positions enhancements with a probability of existence of more than 95% were found. However, the displacements of about $1'$ to the AGN positions make an identification unlikely, an we have to consider the measured photon fluxes as upper limits for the AGN emission. No X-ray source has been found at the position of HB 1304−107, probably due to its very large off-axis angle in the PSPC image. The source image of PG 1411+442 in the hard frequency band (0.6 - 2.4 keV) shows indications of a double structure, which could not be confirmed by the EXSAS maximum likelihood algorithm. However, the the RSDC Standard Analysis Software System (SASS) reports two distinct sources at this position, the stronger one $7''$ from the quasar position and a weaker one about $50''$ in southern direction. We did not include the X-ray bright radio galaxy 3C84 (NGC 1275) in the observation field of 3C83.1, since it is in the center of the Perseus cluster and the X-ray emission is dominated by intracluster gas rather than by the AGN (Böhringer et al. 1993).

The background subtracted number of counts found by the maximum likelihood algorithm, or the 3-$\sigma$ upper limits for uncertain or failed identifications, are given in Tab. 2. The number of counts given for PG 1411+442 has to be understood as the sum of two suspected components, with the stronger one most probably to be identified with the quasar and a contaminating source contributing about 10%.

### 2.3. Spectral analysis

For the spectral analysis, we used the PSPC-AO6 spectral response matrix for the 1991 observations, and the standard Jan94 matrix in all other cases. Raw spectra were determined from photons extracted from a circle of $150''$ around the source, using a background calculated from from a concentric $225''$ to $450''$ annulus around the source. Sources detected by the PSPC image analysis inside this region were cut out with two detection widths (FWHM). In a few cases field sources were found closer than 2-FWHM to the investigated object, and a $90°$ sector centered around their connection line was cut out of the circular field. From the remaining area, source and background photons of the PSPC channels 12 to 240 (about 0.17 to 2.8 keV) were binned requiring a signal-to-noise ratio of at least 4, but increasing that value up to 10 for strong detections to keep the number of data points reasonable.

**Table 2.** Quasars and 3C radio galaxies in the observation fields

| source name | #[a] | net counts | $n_{\mathrm{H(map)}}$[b] $10^{20}$ atoms cm$^{-2}$ | $n_{\mathrm{H(fit)}}$ | $\Gamma$ | photon flux[c] $10^{-4}$cm$^{-2}$s$^{-1}$ | $z$ | $\log L_{[0.5;2]}$ erg s$^{-1}$ |
|---|---|---|---|---|---|---|---|---|
| PG 0026 +129 | 9 | $1094 \pm 33$ | 4.36 | $5.22 \pm 1.05$ | $-2.29 \pm 0.31$ | $45.3 \pm 1.9$ | 0.142 | $44.63 \pm 0.06$ |
| PG 0844 +349 | 2 | $233 \pm 16$ | 3.27 | $3.20 \pm 2.18$ | $-2.40 \pm 0.60$ | $5.6 \pm 0.5$ | 0.064 | $42.87 \pm 0.17$ |
| PG 0953 +414 | 5 | $4676 \pm 71$[d] | 1.14 | $1.46 \pm 0.27$ | $-2.69 \pm 0.13$ | $65.9 \pm 1.1$ | 0.239 | $44.85 \pm 0.05$ |
| PG 1302 −102 | 8 | $1042 \pm 33$ | 3.38 | $2.47 \pm 0.76$ | $-2.10 \pm 0.27$ | $34.3 \pm 1.3$ | 0.286 | $45.00 \pm 0.08$ |
| PG 1411 +442 | 3 | $926 \pm 32$ | 1.13 | $1.18 \pm 0.94$ | $-3.22 \pm 0.57$ | $2.1 \pm 0.1$ | 0.089 | $42.16 \pm 0.30$ |
| PG 1634 +706 | 1 | $544 \pm 24$ | 4.08 | $4.17 \pm 1.55$ | $-2.29 \pm 0.37$ | $6.8 \pm 0.5$ | 1.334 | $45.91 \pm 0.31$ |
| PG 2214 +139 | 7 | $107 \pm 11$ | 4.93 | — | $+0.07 \pm 0.94$[e] | $1.2 \pm 0.3$[e] | 0.066 | $42.38 \pm 0.22$[e] |
| PHL 1657 | 10 | $2456 \pm 50$ | 4.77 | $4.66 \pm 0.65$ | $-2.46 \pm 0.13$ | $90.0 \pm 2.5$ | 0.200 | $45.18 \pm 0.05$ |
| HB 0843 +349 | 2 | $41 \pm 7$ | 3.27 | — | $-2.47 \pm 0.82$[e] | $0.7 \pm 0.2$[e] | 1.575 | $45.05 \pm 0.66$[e] |
| HB 1304 −107 | 8 | $< 22$ | 3.38 | — | $(-2.3)$[f] | $< 1.0$[g] | 2.088 | $< 45.5$[g] |
| HB 2134 −149 | 10 | $< 15$ | 4.77 | — | $(-2.3)$[f] | $< 0.6$[g] | 2.2 | $< 45.4$[g] |
| HB 2135 −145 | 10 | $< 21$ | 4.77 | — | $(-2.3)$[f] | $< 0.7$[g] | 1.9 | $< 45.3$[g] |
| 3C83.1 | 4 | $20 \pm 6$ | 14.54 | — | $(-2.3)$[f] | $0.7 \pm 0.2$[g] | 0.0255 | $41.55 \pm 0.33$[g] |
| 3C433 | 6 | $< 6$ | 11.92 | — | $(-2.3)$[f] | $< 0.7$[g] | 0.102 | $< 42.5$[g] |

[a] observation field number of Tab. 1
[b] general estimated 90% map error: $\pm 10^{20}$atoms cm$^{-2}$ (Elvis et al. 1986)
[c] absorbed flux integrated over the PSPC energy band, using a single-power-law fit with galactic absorption
[d] recalculated from integrated count rate; EXSAS maximum likelihood algorithm limited to 4000 counts
[e] derived for fixed $n_{\mathrm{H(map)}}$, $n_{\mathrm{H}}$ error not considered
[f] estimated value, *not* derived from the data
[g] possible error in $\Gamma$ and $n_{\mathrm{H}}$ not considered

This procedure was performed only for detections with a sufficient number of counts, i.e. for all PG quasars and HB 0843+349. The net source counts used for the spectral analysis may differ from the number of counts given in Tab. 2, in particular it is much lower for sources in crowded fields as PG 1411+442 or HB 0843+349; for the latter one, only 3 spectral data points with a signal to noise ratio of 4 could be extracted. For PG 1411+442, a large number of field sources had to be cut out; furthermore, the 10% contamination by a different X-ray source has to be considered in the following.

In a first step a simple power-law with galactic absorption, treating the galactic $n_{\mathrm{H}}$ as a free parameter, was fitted to the net source spectra. For PG 2214+139 and HB 0843+349, a free fit of $n_{\mathrm{H}}$ was not possible due to low statistics, and we fixed $n_{\mathrm{H}}$ to the value taken from the $n_{\mathrm{H}}$ map used by EXSAS (Dickey & Lockman 1990). For all other PG quasars we found consistency between map and fit values for $n_{\mathrm{H}}$, the error correlation with the spectral index $\Gamma$ is shown in Fig. 1. We calculated the absorbed and unabsorbed photon flux of the sources in the PSPC energy band resulting from the simple power-law fit. The possible error of $n_{\mathrm{H}}$ in the flux error estimates. For the weak observations, it was not possible to fit the spectral index independently, and we calculated their flux by simulating the net count rate (or its upper limit) found by the maximum likelihood method using a power-law spectrum with a fixed index ($\Gamma = -2.3$).

The second section of Tab. 2 shows the map value and the best fit value of the galactic $n_{\mathrm{H}}$ column density ($n_{\mathrm{H(map)}}$ and $n_{\mathrm{H(fit)}}$), the power-law index $\Gamma$ found for best-fit $n_{\mathrm{H}}$, and the absorbed (i.e. observable) photon flux of the sources. Instead of giving an unabsorbed flux, we used the known source redshifts $z$ to calculate the isotropic $0.5 - 2$ keV luminosity $L_{[0.5-2]}$ in the source rest frames, which has the advantage to allow a direct physical comparison. The given errors represent approximately the 1-$\sigma$ confidence intervals for strong detections. The fluxes and luminosities given for PG 2214+139, HB 0843+349 and 3C83.1, and the 3-$\sigma$ upper limits given for uncertain detections, have to be understood as order of magnitude estimates rather, since not all possible errors have been considered. The high-redshift quasar HB 0843+349 has a luminosity and spectral index comparable to the strong PG quasars. PG 2214+139 seems to have an unusually flat spectrum, possibly indicating internal absorption on neutral hydrogen. The luminosity of 3C83.1 is about one order of magnitude below that of our weakest quasars, suggesting strong absorption of X-rays emitted from the active nucleus, as expected in unified AGN models (Falcke et al. 1995a).

### 2.4. Model fits

The detailed modeling of quasar soft-X-ray spectra discussed in the following is performed only for the seven strong de-tections of our sample. We compare the quality of spectral fits obtained with a single-power-law model (*1-pl*) with two different two-component models: (i) a thermal-plus-power-law model (*th+pl*) characterized by a maximum temperature and a hard-X-ray power-law slope (for the sake of comparison identical to that used by Walter et al. 1994), and (ii) a two-power-law model (*2-pl*) characterized by two power-law slopes. To keep the number of free parameters reasonable, we fixed the galactic $n_H$ to the best available value: In four of seven observations the best-fit and map values of $n_H$ correspond much better than expected from the error estimates, and we used the map values only. In two cases, PG 0026+129 and PG 1302−706, the values hardly agree within their 90% error bars, and we perform the analysis for both. For PG 0953+414 the map value lies outside the 2-$\sigma$ error regime of the best-fit value, but the map error is so large that we can expect that the best-fit value derived from a single-power-law fit provides a better estimate of the true hydrogen column density, and we used it henceforth. We fixed the boundary conditions for the two-component models in the following way: For the photon spectral index of the hard component we used $\Gamma_{hx} = -2$ for radio-quiet AGN and $\Gamma_{hx} = -1.5$ for radio-loud AGN (Wilkes & Elvis 1987, Brunner et al. 1994); for PG 0026+129 we took the *EXOSAT* value $\Gamma_{hx} = -1.86$ (Comastri et al. 1992), for PHL 1657 the *Ginga* value $\Gamma_{hx} = -1.81$ (Williams et al. 1992). The flux of the soft component was fixed to the observed UV flux at an energy of 10 eV in the *source frame*, $f_{0.01}$, corrected for galactic absorption. The soft component of the *th+pl* model was designed, in correspondence to Walter et al. (1994), as a power-law spectrum with exponential cutoff, $f_E^{uv} \propto E^{-1.3} \exp(-E/T_c)$. The free fit parameters are $\Gamma_{\rm sx}$ and $T_c$ for the *2-pl* and *th+pl* model, respectively, and the flux of the hard-X-ray component at 2 keV, $f_2^{\rm hx}$, in both models. A schematic comparison of the shape of both models in the UV-to-X frequency range is shown in Fig. 2. It is obvious from this diagram, that the difference of these two models and the single power-law model *which can be tested by our observations* is their "curvature" in the PSPC energy band.

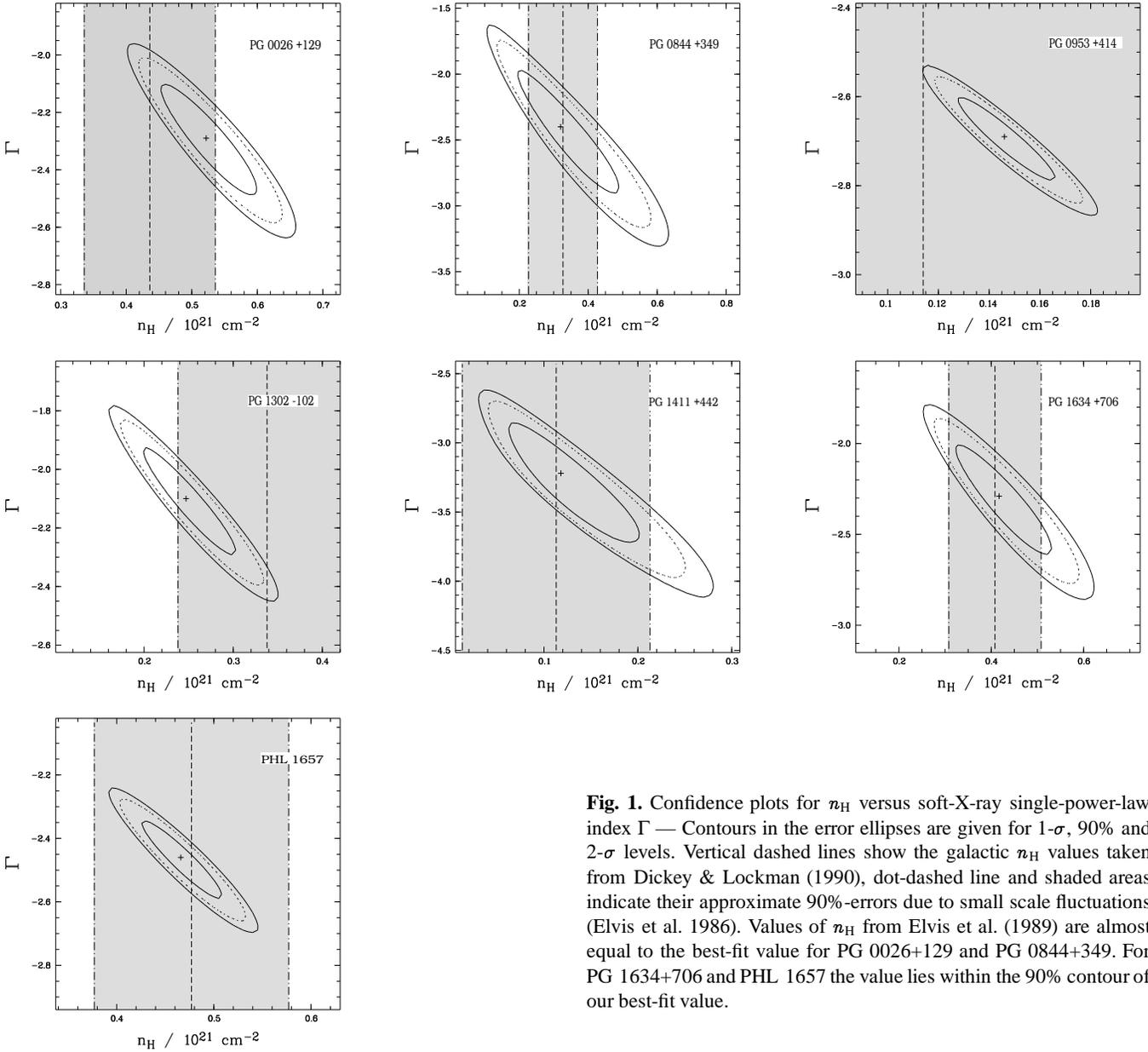

**Fig. 1.** Confidence plots for $n_H$ versus soft-X-ray single-power-law index $\Gamma$ — Contours in the error ellipses are given for 1-$\sigma$, 90% and 2-$\sigma$ levels. Vertical dashed lines show the galactic $n_H$ values taken from Dickey & Lockman (1990), dot-dashed line and shaded areas indicate their approximate 90%-errors due to small scale fluctuations (Elvis et al. 1986). Values of $n_H$ from Elvis et al. (1989) are almost equal to the best-fit value for PG 0026+129 and PG 0844+349. For PG 1634+706 and PHL 1657 the value lies within the 90% contour of our best-fit value.

**Table 3.** Comparison of the *1-pl*, *2-pl* and *th+pl* model fits to the PSPC data for six strong PG quasars and PHL 1657 — $n$ is the number of data points, the degrees freedom are $\nu = n - 2$ in all models. $\chi^2_\nu \equiv \chi^2/\nu$ expresses the fit quality, $z_b$ the significance for an excess-like correlation of the residuals in standard deviations (see text and Fig. 2). $n_{\rm H}$ is the galactic column density of neutral hydrogen in $10^{20}$ atoms cm$^{-2}$, $f_{0.01}$ the dereddened photon flux at 10 eV, $f_2^{\rm hx}$ the observed best fit flux of the *hard-X-ray component* at 2 keV; photon fluxes are given in keV$^{-1}$ cm$^{-2}$ sec$^{-1}$, photon energies refer to the source frame. $\Gamma_{\rm hx}$ and $\Gamma_{\rm sx}$ are the hard- and soft-X-ray spectral indices, respectively, the latter defined in the *2-pl* model only, $T_c$ the cutoff temperature in the *th+pl* model, given in eV. When model fits are performed on the same data for different boundary conditions, a tilde ($\sim$) indicates that a value is unaffected by the differences. The UV flux marked with * for PHL 1657 is constructed from a best *2-pl* fit for fixed $f_2^{\rm uv}$, the errors of $f_{0.01}$ and $\Gamma_{\rm sx}$ are not comparable with other fits.

| source name | data $n$ | fixed model parameters $n_{\rm H}$ | $f_{0.01}$ | $\Gamma_{\rm hx}$ | *1-pl* model $\Gamma$ | $\chi^2_\nu$ | $z_b$ | *2-pl* model $10^5 f_2^{\rm hx}$ | | $\Gamma_{\rm sx}$ | $\chi^2_\nu$ | $z_b$ | *th+pl* model $T_c$ | $10^5 f_2^{\rm hx}$ | $\chi^2_\nu$ | $z_b$ |
|---|---|---|---|---|---|---|---|---|---|---|---|---|---|---|---|---|
| PG 0026+129 | 21 | 4.36 | 287±56 | −1.86 | −2.08±0.08 | 0.7 | −1.6 | 52.6±5.3 | | −3.05±0.14 | 0.9 | −3.1 | 41.9±4.4 | 58.7±3.0 | 1.2 | −5.0 |
| PG 0026+129 | ∼ | 5.22 | 320±64 | ∼ | −2.29±0.09 | 0.5 | +0.3 | 37.8±9.0 | | −2.78±0.08 | 0.6 | −0.8 | 51.6±2.8 | 59.9±3.1 | 1.3 | −5.6 |
| PG 0844+349 | 8 | 3.27 | 253±55 | −2.0 | −2.42±0.19 | 1.2 | −0.3 | 4.1±1.0 | | −3.50±0.16 | 1.4 | −1.0 | 32.7±2.6 | 4.7±0.7 | 1.6 | −2.3 |
| PG 0953+414 | 34 | 1.46 | 394±60 | −2.0 | −2.69±0.04 | 0.8 | +0.5 | 3.7±3.2 | | −2.75±0.02 | 0.8 | −0.4 | 57.2±0.7 | 30.9±1.3 | 3.8 | −15. |
| PG 1302−102 | 22 | 3.38 | 356±70 | −1.5 | −2.39±0.08 | 1.0 | +2.0 | 21.3±5.0 | | −2.79±0.03 | 0.9 | +0.3 | 56.8±1.3 | 43.9±2.8 | 2.1 | −7.3 |
| PG 1302−102 | ∼ | 2.47 | 317±56 | ∼ | −2.10±0.07 | 0.8 | −0.1 | 33.5±3.7 | | −2.96±0.04 | 1.1 | −2.6 | 48.8±1.4 | 43.4±2.7 | 2.1 | −8.4 |
| PG 1411+442 | 13 | 1.13 | 106±35 | −2.0 | −3.19±0.20 | 0.8 | −0.1 | 0.2±0.1 | | −3.50±0.04 | 0.7 | −0.6 | 29.7±0.7 | 0.5±0.1 | 2.7 | −7.6 |
| PG 1634+706 | 14 | 4.08 | 703±122 | −2.0 | −2.27±0.14 | 0.7 | −0.7 | 12.6±1.9 | | −3.30±0.14 | 0.7 | −1.7 | 64.4±5.6 | 14.3±1.2 | 1.0 | −3.6 |
| PHL 1657 | 27 | 4.77 | 84±31 | −1.81 | −2.50±0.05 | 0.8 | +1.1 | 50.0±38. | | −2.26±0.12 | 5.3 | −2.4 | 104.±3.7 | 105.±4.0 | 3.0 | −11. |
| PHL 1657 | ∼ | ∼ | 1165±385* | ∼ | ∼ | ∼ | ∼ | 50.0±0.0* | | −2.84±0.08* | 0.8 | −0.8 | 52.3±1.1 | 106.±3.9 | 3.6 | −11. |

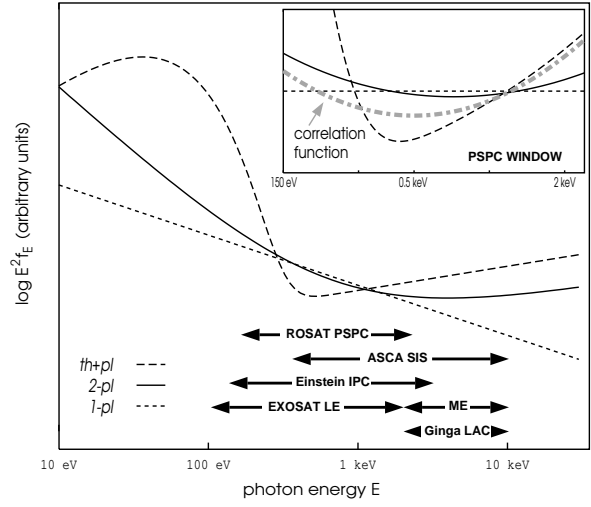

**Fig. 2.** Schematic comparison of the fit models used in this paper — The main energetic differences occur in the unobservable EUV range, which is shown here in a broad band plot of $E^2 f_E$ vs. photon energy (log-log). The sensitivity ranges of various X-ray satellite spectrometers are indicated. The insert frame shows the situation in the PSPC band: $f_E/E^\Gamma$ (linear) is plotted vs. photon energy (log), and may be compared with the residual plots in Fig. 3. Also shown is the correlation function used to calculate the "curvature index" $z_b$ (see text).

Table 3 shows the best fit parameters of the various model fits. For the 2-pl model we find a mean soft-X-ray slope of $\Gamma = -3.1 \pm 0.3$, for the *th+pl* model a mean temperature of $55 \pm 23$ eV; where two different fits were made for the same source, only the better one (lower $\chi^2$) was used for averaging. The UV/hard-X-ray flux ratio scatters over a broad range, $5.7 < \log(f_{0.01}/f_2^{\rm hx}) < 7.7$, without any obvious correlation to other source properties. For PG 0953+414 no clear evidence for the existence of a hard component can be drawn from the PSPC data, since the single-power-law spectrum alone connects to the UV flux, with a power-law index $\Gamma$ in good correspondence with the *soft component* indices $\Gamma_{\rm sx}$ found for other sources. For PG 0026+129 the single power-law index is consistent with the *EXOSAT* value within the observational errors, hence no clear evidence for the existence of a soft-X-ray excess in the PSPC band is given. To connect to the UV flux, however, a steepening of the spectrum in the EUV is necessary, again yielding a soft power-law index comparable to that of other sources. For the weak-bump quasar PHL 1657 the observed UV flux is much too low to allow a *2-pl* fit, but nevertheless the source shows a soft-X-ray excess which has the same general shape as for the big-bump PG quasars. Using the PSPC data to construct a best-fit UV flux for the *2-pl* model by minimizing $\chi^2$ for fixed hard-X-ray properties, yields a UV/hard-X-ray flux ratio comparable to that found for the strong bump quasars.

Figures 3 and 4 show the residuals of the model fits. For most sources, it is obvious that the residuals for *th+pl* show a systematic deviation, which is absent in the *1-pl* and *2-pl* models, indicating that the curvature, i.e. the spectral break, of the *th+pl* model in the PSPC band systematically exceeds the observed value. For the discussion of the fit quality it therefore seems to be helpful to quantify the "curvature-effect" expressing the specific differences of our fit models. One obvious way do this is to calculate the correlation coefficient $r_b$ of the residuals $R(x_i)$ to $b(x_i) = x_i^2$, with $x_i = \log(E_i/E_c)$, and $E_c = 0.5\,\mathrm{keV}$ defining the logarithmic center of the PSPC band. Then, the index

$$z_b = \frac{r_b\sqrt{n}}{1 - r_b^2} \qquad (1)$$

expresses the significance of a quadratic correlation of the residuals in terms of standard deviations. From Tab. 3 it becomes obvious that $z_b$ allows a significant distinction between the models even where $\chi^2$ remains reasonably low in all cases (e.g. PG 0026+129), and expresses very well the impact drawn from inspection of the residual plots. Another merit of the *signed* quantity $z_b$ shows up in the discussion of the effect of the assumed value for the galactic $n_\mathrm{H}$, which also can cause a curvature effect due to energy dependent absorption: We see from comparison of the fits made for the same sources at different $n_\mathrm{H}$, that the typical differences between the *1-pl* and *2-pl* models can be removed by varying $n_\mathrm{H}$ within the range allowed by the uncertainties of the galactic map, with $z_b$ increasing with $n_\mathrm{H}$. In particular, for PG 1302−102 the map value for $n_\mathrm{H}$ gives better results for the *2-pl* model than the *1-pl* best-fit $n_\mathrm{H}$. We find that the galactic $n_\mathrm{H}$ error only allows a change of $\Delta z_b \lesssim 3$, but we have $\Delta z_b > 3$ between *1-pl* and *th+pl* for most of our sources. This may give rise to the assumption that the large negative $z_b$ values of the th+pl model may be corrected by assuming strong intrinsic absorption in the source. Fitting the data of PG 0953+414 with a *th+pl* model for free $n_\mathrm{H}$ gives a minimal $\chi^2$ for $n_\mathrm{H} = 2.54 \cdot 10^{20}$ atoms cm$^{-2}$, but in fact only the data below 0.5 keV can be flattened to the power-law slope this way; the remaining curvature index $z_b = -3.3$ and $\chi_\nu^2 = 2.4$ show that the correction by intrinsic absorption is insufficient to produce an acceptable fit.

Finally we are led to the following conclusion: The difference between *the sharp spectral break expected for a Wien-tailed UV-to-soft-X-ray excess and a simple power-law is detectable* in the PSPC range and cannot be removed by variations of the neutral hydrogen absorption strength, while a *slight curvature expected for a superposition of two power-laws is indistinguishable from a single power-law*. Clearly, this applies to any other model expecting a weakly concave X-ray spectrum as well.

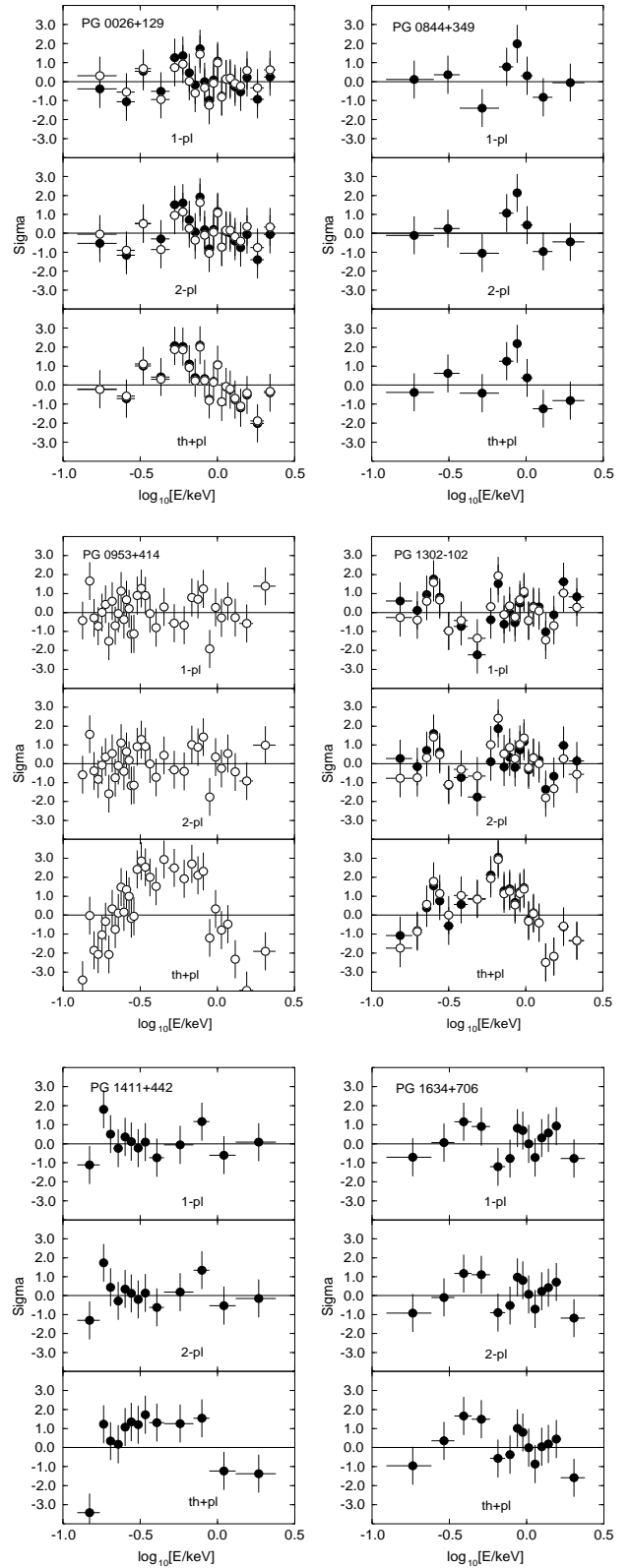

**Fig. 3.** Residuals for single-power-law, two-power-law and thermal+power-law model fits (standard deviations vs. photon energy) — Filled circles correspond to the galactic map value, open circles to the best fit value of $n_\mathrm{H}$.

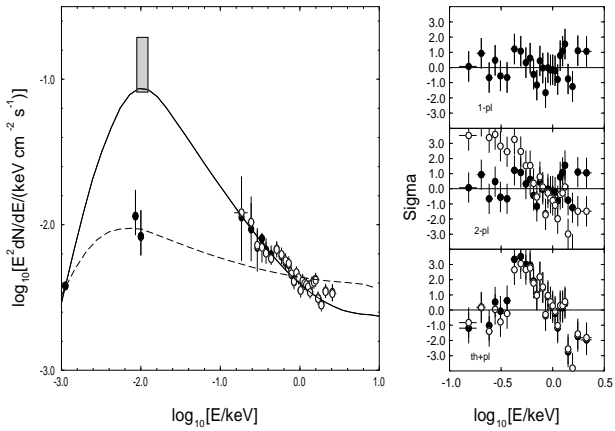

**Fig. 4. Left:** Spectral energy distribution of the "weak-bump" quasar PHL 1657 in the source frame — The optical data point was taken from McDowell et al. (1989); UV fluxes were taken from Cheng & Fang (1987) and extracted from Lanzetta et al. (1993). The latter value was dereddened according to Seaton (1979). The hard-X-ray slope was set to the value measured by Williams et al. (1992) with *Ginga*. The solid line shows a theoretical model spectrum connecting to a much higher intrinsic UV flux at the source (shaded area) than observed by the IUE, the dashed curve represent a model connection to the the observed weak UV flux (see Mannheim et al. 1995 for details, and Sect. 3.4. for a discussion). Data are binned accordingly for both models (filled and open circles, respectively). **Right:** Residuals of various model fits to the X-ray spectrum as described in Sect. 2.4. — In the X-ray regime, the two-power-law model is equivalent to the theoretical spectrum.

## 3. Discussion

### 3.1. The shape of the soft-X-ray spectrum

The general result for our small heterogeneous sample is that the source spectra are best represented by structureless power-laws with a mean slope of $\Gamma = -2.5 \pm 0.4$, adopting the galactic map value of the neutral hydrogen column density. This is in reasonable agreement with $\Gamma = -2.7 \pm 0.04$ obtained for the mean spectrum of a number of low-redshift quasars observed with high signal-to-noise ratio by the *ROSAT* PSPC (Fiore et al. 1994). These power-laws are systematically steeper than the power-laws found in the hard-X-ray range by other satellite experiments by $\Delta\Gamma = 0.5\text{-}0.7$ (e.g., Wilkes & Elvis 1987, Brunner et al. 1994, Jackson et al. 1993, Fiore et al. 1994). Fiore et al. 1994 have concluded from a thorough analysis of the PSPC/IPC cross-calibration, that the systematic differences between both instruments are larger than the systematic uncertainty of the PSPC spectral index. Further constraints on the significance of this result will be provided in the near future using results of the cross-calibration of the PSPC with the high resolution X-ray spectrometers SIS and GIS on board of the Advanced Satellite for Cosmology and Astrophysics (*ASCA*); a recent comparison of high signal-to-noise *ASCA* and *ROSAT* spectra of PKS 0537–286 indicates consistency of the single-power-law indices (Siebert et al. 1995).

Our results are in agreement with Fiore et al. (1994), in that the soft-X-ray spectra are dominated by a single *broad* component, which argues against a Wien-shaped spectrum connecting soft X-rays and UV bump. The latter possibility has been advocated by Walter & Fink (1993) and Walter et al. (1994) from a study of a large sample of AGN including low-luminosity Seyferts, implying a very high effective temperature of the putative thermal emission in AGN of the order $T_{\mathrm{eff}} \sim 50$ eV. To facilitate a comparison with our quasar data, we repeated their analysis. As a result, we also obtained $T_{\mathrm{eff}} \sim 50$ eV, but the corresponding $\chi^2$ values are unacceptably large due to a systematical excess of photons in the energy range about $\sim 0.5$ keV. The disagreement becomes most apparent for the spectra with highest signal-to-noise, and cannot be completely removed by the assumption of intrinsic absorption. Moreover, a highly curved soft X-ray spectrum would quickly be shifted out of the PSPC bandpass for higher redshifts. In agreement with Puchnarewicz et al. (1994), we see no clear evidence for this; the disappearance of the soft component seems to occur only gradually with redshift. Again, we come to the conclusion that a single *broad* component dominates the soft X-rays in quasars.

The broad soft-X-ray component is constrained by the requirement that it must not overproduce the UV photons and connect to the hard-X-ray spectrum. We therefore considered the simplest possibility, i.e. a two-power-law spectrum with one spectrum starting in the UV at the dereddened flux measured by IUE and the other spectrum with a typical or measured hard-X-ray slope. For the soft-X-ray power-law we obtained a mean slope of $\Gamma_{\mathrm{sx}} = -3.1 \pm 0.3$ in agreement with the value $\Gamma_{\mathrm{sx}} = -3.3 \pm 0.3$ found by Fiore et al. (1994). The latter value was obtained without enforcing that the the soft-X-ray power-law starts at the observed UV flux. However, we found from our analysis that the parameter sets with lowest $\chi^2$ values, in fact, also give the UV fluxes in the correct range. Physically, a two-power-law spectrum is very appealing: (i) The need for a higher central temperature is removed, (ii) bolometric corrections due to the soft X-rays are small and (iii) there is a natural radiation process to account for such a spectrum, viz. unsaturated Comptonization by some sort of coronal high temperature gas (Titarchuk 1994). Most remarkably, recent EUV measurements have brought unambiguous evidence for an AGN with a soft X-ray power-law spectrum (Wisotzki et al. 1995).

### 3.2. Reprocessing of primary X-rays

On the basis of *ROSAT* observations alone, it is barely possible to distinguish two power-laws from a blend of emission lines, absorption edges and a power-law. For some Seyferts, partially ionized material seems to imprint signatures on the soft-X-ray spectra either in the line of sight (Yaqoob & Warwick 1991, Turner et al. 1993, Pounds et al. 1994) or, possibly, from a reflecting slab (Czerny & Życki 1994). Such features can mimic a steep soft-X-ray spectrum when folded with the response of a detector with poor energy resolution. However, the observations still require the existence of an additional steep soft-X-ray spectral component in many sources. Moreover, if there is a uni-versal hard-X-ray power-law slope $\Gamma_{\rm hx} \simeq -2$ for radio-quiet AGN, and if the observed value $\Gamma_{\rm hx} \simeq -1.7$ for Seyferts is due to reflection (Pounds et al. 1990), then reprocessing should be comparatively unimportant in quasars. Hence, soft excesses would not be expected, contrary to the observations. We also note that the observed rather moderate dispersion of the PSPC soft-X-ray slope observed by us would not be expected in a random sample of PG quasars. Other arguments against reprocessing by partially ionized matter as the general origin of soft-X-ray excesses are the extreme steepness of the X-ray spectra observed in *ROSAT*-selected AGN (Grupe et al. 1995), the rapid soft-X-ray variability (Boller et al. 1993, Molendi & Maccacaro 1994), and the general shape of spectra observed over a broader X-ray range (Fiore et al. 1994, Yaqoob et al. 1994).

### 3.3. Soft X-rays and accretion disks

A soft-X-ray spectrum connecting to the thermal UV-bump spectrum of an accretion disk requires hot plasma $T_{\rm e} > T_{\rm eff}$ in the immediate vicinity of the cold disk. Unsaturated Compton-scattering of disk photons produces a soft-X-ray power-law. Major scenarios for the origin of the hot electron plasma are:

1. Accretion at a near-Eddington rate. This generates an electron scattering corona above the disk. For a sufficiently large optical depth a Wien-shaped extension of the UV-bump reaching up to the soft-X-ray range results (Ross & Fabian 1993).

2. A two-temperature domain of the accretion disk (Wandel & Liang 1991), as has been proposed for Cyg X-1 (Shapiro et al. 1976). Due to secular and thermal instabilities the disk blows up to a torus sustained by the pressure of protons with $10^9$ K, whereas the electrons remain much colder. Comptonization of photons from the outer, colder parts of the disk traversing the torus leads to the soft X-rays. A variant of this idea has been proposed recently by Narayan et al. (1995) in the context of an advection dominated disk model. Stable solutions with a hot central region and a cold outer disk exist in this model.

3. A disk/corona system where most energy dissipation occurs in the corona instead of in the disk (Haardt & Maraschi 1991). This would generally produce very hot material with sufficient optical depth above the cold disk producing Comptonized soft *and* hard X-rays.

4. Accretion onto a Kerr black hole (Dörrer et al. 1995). The superposition of Comptonized photon spectra from an accretion disk with vertical temperature gradient around a maximal rotating black hole leads to non-Wien-shaped big blue bump extending up to about 3 keV. In equatorial directions, relativistic blueshifting extends the observed spectrum to even higher energies, with a large fraction of the total energy emitted as soft and hard X-rays.

5. The optically thick base of a disk wind (Mannheim et al. 1995). To obtain such a medium the disk must suffer severe mass loss which is not possible for a radiation-driven wind. MHD winds can provide such mass losses and, at the same time, transport away the angular momentum of the accretion flow, thereby explaining the large accretion rates required to obtain high quasar luminosities. The mechanism works, if open poloidal magnetic field lines are anchored to the disk (e.g., Ferreira & Pelletier 1993).

Scenario 1 implies steep Wien-shaped X-ray extensions of a thermal UV spectrum as suggested by Walter & Fink (1993) and Walter et al. (1994). However, such spectra yield a very poor representation of our data (Sect. 2.5). Scenario 2 seems physically problematic, since it is not clear whether one can decouple the proton from the electron temperature by many orders of magnitude in a steady state, so that the gas pressure becomes large enough to dominate the disk structure. In the absence of a confining pressure it is not possible to find boundary conditions corresponding to hydrostatic equilibrium (Parker 1963). Scenario 3 is attractive from a phenomenological point of view, but there is no self-consistent calculation of the disk structure for this kind of energy dissipation law. Scenario 4 predicts slightly flattening UV-to-X-ray spectra, in qualitative agreement with our data. However, for the model expects a cutoff in the 2–10 keV regime, which is not observed in quasars; an additional hard-X-ray source is thus required. On the other hand, it provides an appealing explanation for strong soft-X-ray emission in weak-bump quasars (see below). Scenario 5 depends on the hypothesis that disk winds with luminosities of the same order as the electromagnetic disk luminosity are required. There is some support for this hypothesis from (a) a unification scenario of AGN (Falcke et al. 1995b) (b) the ability of the disk wind to produce nonthermal hard X-rays extending over the whole sensitivity range of actual X-ray satellites (Mannheim 1995) (c) the prediction of $\Gamma_{\rm sx} \sim -3$ for the soft-X-ray slope and $\epsilon_{\rm b} \sim 1$ keV for the break to the hard X-rays (Mannheim et al. 1995).

Detailed comparison with high signal-to-noise quasar spectra are required to test the predictivity of these models and discern their specific differences. In general, most models assuming saturated Comptonization lead to nearly Wien-shaped soft-X-ray spectra, which are too steep to be in agreement with the data. Unsaturated Comptonization alone is insufficient to explain the hard-X-ray spectra, strengthening our belief that one really needs a nonthermal process in addition to the thermal processes producing the soft X-ray emission.

### 3.4. A soft-X-ray excess in a weak-bump quasar?

The weak-bump quasar PHL 1657 exhibits the same type of soft-X-ray power-law as the PG quasars of our sample. PHL 1657 is known to exhibit large-amplitude continuum variability (Smith et al. 1993) which, if considered as an indication of a variable strength of the big blue bump, could explain the observations. Another possibility is internal reddening in an ionized medium containing large amounts of dust. Unusual properties are indicated, indeed, by the Balmer decrement $I(H_\alpha)/I(H_\beta) = 17$ (Zheng et al. 1987). Moreover, PHL 1657 is a near-symmetric FR-II radio source (Morganti et al. 1993) and the flux ratio of extended to core radio emission argues for a large inclination angle of the jet axis to the observer, thus the low inclination angle of the putative accretion disk leads to higher intervening column densities. From the X-ray observations, however, we did not find evidence for much cold gas.

An alternative explanation could be given by assuming accretion onto a Kerr black hole, in which case the peak of the energy emission is shifted to the unobservable EUV band. The results of Dörrer et al. 1995 for a maximally rotating black hole, seen under a large angle to the accretion disk axis (equivalent to the jet axis) seem to be able to explain the spectrum of PHL 1657 without calling for intrinsic absorption.

Clearly, high-resolution X-ray spectra, which could be obtained by *ASCA* or the future X-ray observatories *AXAF* or *XMM*, are required to discern the primary continuum spectrum of this exceptional source. Moreover, observations of other weak-bump quasars could give important information for our understanding of the origin of soft X-rays in AGN.

## 4. Conclusions

As a typical feature observed in a small, randomly selected quasar sample we find that the PSPC soft-X-ray data consistently fit into the picture of an overall, weakly concave spectral shape. If interpreted as two superimposed power-law spectra, the soft X-ray component with slope $\Gamma_{sx} = -3.1 \pm 0.3$ extrapolates down to meet the UV flux at about $\sim 10\,\mathrm{eV}$. A natural interpretation of these findings is that the soft X-rays in quasars are of unsaturated Comptonization origin, covered by a flat, nonthermal hard-X-ray component above $\epsilon_b = 1.4 \pm 1.1\,\mathrm{keV}$.

The strength of the hard-X-ray component compared to the UV emission varies drastically among the quasars of our sample, but it is found that the observed X-rays are always energetically unimportant. An exception to this could be PHL 1657, which does not appear as a weak-bump quasar in soft X-rays. It has the same soft X-ray spectrum as the strong-bump PG quasars. Possible reasons for this are internal reddening by an ionized medium containing significant amounts of dust, a variable bump, or relativistic blueshifting of the spectrum emitted by accretion onto a Kerr black hole.

*Acknowledgements.* U. Zimmermann is gratefully acknowledged for helpful advice. We also acknowledge the referee H. Brunner for his remarks. This research was supported by DARA grant FKZ 50 OR 9202. The NASA/IPAC Extragalactic Database (NED) used for this research is operated by the Jet Propulsion Laboratory, California Institute of Technology, under contract with the NASA.

## References

Bechtold J., Green R.F., Weymann R.J., et al., 1984, ApJ 281, 76
Böhringer H., Voges W., Fabian A.C., et al., 1993, MNRAS 264, L25
Boller T., Trümper J., Molendi S., et al., 1993, AA 279, 53
Brunner H., Lamer G., Worrall D.M., Staubert, R., AA 287, 436
Cheng F.H., Fang L.Z., 1987, MNRAS 226, 485
Chini R., Kreysa E., Biermann P.L., 1989, AA 219, 87
Clavel J., Nandra K., Makino F., et al., 1992, ApJ 393, 113
Comastri A., Setti G., Zamorani G., et al., 1992, ApJ 384, 62
Czerny B., Elvis M., 1987, ApJ 321, 305
Czerny B., Życki P.T., 1994, ApJ 431, L5
Dickey J.M., Lockman F.J., 1990, ARAA 28, 215
Dörrer T., Riffert H., Staubert R., Ruder H., 1996, AA in press
Elvis M., Green R.F., Bechtold J., et al., 1986, ApJ 310, 291
Elvis M., Lockman F.J., Wilkes B.J., 1989, AJ 97, 777
Falcke H., Gopal-Krishna, Biermann P.L., 1995a, AA 298, 395
Falcke H., Malkan M., Biermann P.L., 1995b, AA 298, 375
Fanaroff B.L., Riley J.M., 1974, MNRAS 167, 31P
Ferreira J., Pelletier G., 1993, AA 276, 625
Fiore F., Elvis M., McDowell J.C. et al., 1994, ApJ 431, 515
Grupe D., Beuermann K., Mannheim K., et al., 1995, AA 300, L21
Guilbert P.W., Rees M.J., 1988, MNRAS 233, 475
Haardt, F., Maraschi, L., 1991, ApJ 380, L51
Hewitt A., Burbidge G., 1991, ApJS 75, 451
Jackson N., Browne I.W.A., Warwick R.S., AA 274, 79
Kellermann K.I., Sramek R., Schmidt M., et al., 1989, AJ 98, 1195
Lanzetta K.M., Turnshek D.A., Sandoval J., 1993, ApJS 84, 109
Laor A., Netzer H., 1989, MNRAS 238, 897
Lightman A.P., White T.R., 1988, ApJ 335, 57
Madau P., 1988, ApJ 327, 116
Maisack M., Collmar W., Barr P. et al., 1995, AA 298, 400
Mannheim K., Biermann P.L., 1992, AA 253, L21
Mannheim K., 1995, AA 297, 321
Mannheim K., Schulte M., Rachen J.P., 1995, AA 303, L41
Maraschi L., Ghisellini G., Celotti A., 1992, ApJ 397, 5
Molendi S., Maccacaro T., 1994, AA 291, 420
McDowell J.C., Elvis M., Wilkes B.J. et al., 1989, ApJ 345, L13
Morganti R., Killeen N.E.B., Tadhunter C.N., MNRAS 263, 999
Narayan R., 1995, ApJ 444, 231
Netzer H., 1993, ApJ 411, 594
Niemeyer M., 1995, PhD thesis, MPI f. Radioastron., Bonn, Germany
Parker, E.N., "Interplanetary Dynamical Processes", Wiley and Sons, New York-London, 1963
Pfeffermann E., et al., 1987, Proc. SPIE 733, 519
Pounds K.A., Nandra K., Stewart G.C., et al., 1990, Nat 344, 132
Pounds K.A., Nandra K., Fink H.H., Makino F., 1994, MNRAS 267, 193
Puchnarewicz E., Mason K.O., Córdova F.A., 1994, MNRAS 270, 663
Rees M.J., 1984, ARAA 22, 471
Reimers D., Vogel S., Hagen H.-J., et al., 1992, Nat 360, 561
Reynolds C.S., Fabian A.C., 1995, MNRAS 273, 1167
Ross R.R., Fabian A.C., Mineshige S., 1992, MNRAS 258, 189
Ross R.R., Fabian A.C., 1993, MNRAS 261, 74
Saxton R.D., Turner M.J.L., Williams O.R., et al., 1993, MNRAS 262, 63
Seaton M.J., 1979, MNRAS 187, 73P
Shakura N.I., Sunyaev R.A., 1973, AA 24, 337
Shapiro S.L., Lightman A.P., Eardley D.M., 1976, ApJ 204, 187
Siebert J., Matsuoka M., Brinkmann W., et al., 1995, AA in press
Smith A.G., Nair A.D., Leacock R.J., Clements S.D., 1993, AJ 105, 437
Sun W.-H., Malkan M.A., 1989, ApJ 346, 68
Titarchuk L., 1994, ApJ 434, 570
Trümper J., 1983, Adv. Space Res. 2, No.4, 241
Turner T.J., George I.M., Mushotzky R.F., 1993, ApJ 412, 72
Walter R., Fink H.H., 1993, AA 274, 105
Walter R., Orr A., Courvoisier T.J.-L., 1994, AA 285, 119
Wandel A., 1991, AA 241, 5
Wandel A., Liang E.P., 1991, ApJ 380, 84
Wilkes B.J., Elvis M., 1987, ApJ 323, 243
Williams O.R., Turner M.J.L., Stewart G.C., et al., 1992, ApJ 389, 157
Wisotzki L., et al., 1995, AA 297, L55
Yaqoob T., Warwick R.S., 1991, MNRAS 248, 773
Yaqoob T., Serlemitsos P., Mushotzky R., et al., 1994, PASJ 46, L173
Zdziarski A.A., Lightman A.P., 1985, ApJ 294, L79
Zheng W., Burbidge E.M., Smith H.E., et al., 1987, ApJ 322, 164
Zimmermann H.U., et al., 1994, EXSAS User's Guide, MPE Rep. 257, Max-Planck-Institut für extraterr. Physik, Garching, Germany.